\documentclass[aps,showpacs,preprint,amsmath,amssymb,pra]{revtex4-1}

\usepackage{graphicx} 
\usepackage{color} 
\usepackage{bm} 
\usepackage{acronym}

\usepackage{amssymb} 

\usepackage{bbold} 
\usepackage{braket}

\newcommand{\real}{\rm I\!R} 
\newcommand{\imag}{\rm I\!I}

\usepackage{xr}

\begin{document} 

\title{Inverse-designed metaphotonics for hypersensitive detection}

\author{Maxim S. Elizarov$^1$, Yuri S. Kivshar$^{2,3}$, and A. Fratalocchi$^{1}$} 
\date{\today} 
\email{andrea.fratalocchi@kaust.edu.sa} 
\homepage{www.primalight.org}
\affiliation{
  $^1$PRIMALIGHT, Faculty of Electrical Engineering; Applied Mathematics and Computational Science, KAUST, Thuwal 23955-6900, Saudi Arabia}
  \affiliation{
  $^2$Australian National University, Canberra ACT 2601, Australia } 
 \affiliation{$^3$ITMO University, St.~Petersburg 197101, Russia}

\begin{abstract}
Controlling the flow of broadband electromagnetic energy at the nanoscale remains a critical challenge in optoelectronics. Surface plasmon polaritons (or plasmons) provide subwavelength localization of light, but are  affected by significant losses. On the contrary, dielectrics lack a sufficiently robust response in the visible to trap photons similar to metallic structures. Overcoming these limitations appears elusive, as it implies devising a path to circumvent causality in the quantum-mechanical form of matter. Here we demonstrate that addressing this problem is possible if we employ a novel approach based on suitably deformed reflective metaphotonic structures. The complex geometrical shape engineered in these reflectors emulates nondispersive index responses, which can be inverse-designed following arbitrary form factors. We discuss the realization of essential components such as resonators with an ultra-high refractive index of $n=100$ in diverse profiles. These structures support localization of light in the form of bound states in the continuum (BIC), fully localized in air, in a platform in which all refractive index regions are physically accessible. We discuss our approach to sensing applications, designing a class of sensors where the analyte directly contacts areas of ultra-high refractive index. Leveraging this feature, we report differential sensitivities up to $350$~nm/RIU in structures with footprints of approximately one micron. These performances are two times better than the closest competitor with a similar form factor. Inversely designed reflective metaphotonics offers a flexible technology for controlling broadband light, supporting optoelectronics' integration with large bandwidths in circuitry with miniaturized footprints.
\end{abstract} 

\maketitle

\section*{Introduction}
In applications that exploit the propagation of electromagnetic beams, it is crucial to control broadband fields in small volumes of matter. At GHz frequencies, this reduces chip footprint and increases
device speed. At optical wavelengths, strongly localized fields provide the key to many significant effects, including low-threshold lasing, efficient nonlinear harmonic generations, high-resolution imaging, highly sensitive detection, low-power communications, enhanced security, and complex wavefront engineering~\cite{lee2019ultrahigh,koshelev2019nonradiating,Koshelev288,wu2021bound,huttenhofer2021anapole,wang_maier_crescent_met_2021,Nasir_Zayats_2021,smirnova2020nonlinear,chen2021topologically}. However, despite impressive recent progress in this field, photonics integration still lags behind electronics, which employs now pioneering devices with characteristic scales of 10~nm or less~\cite{el0,el1}.\\
To trap light at subwavelength scales, one employs a traditional approach based on collective excitations of electrons and electromagnetic waves in metallic resonators supporting hybrid modes described by surface plasmon polaritons. Currently, metals provide the highest degree of light confinement via plasmonic resonance~\cite{maier}. However, substantial losses at optical wavelengths create a significant challenge to exploiting these materials for advanced applications~\cite{franano,dar0,koshelev2019nonradiating}.\\
The recently emerged field of dielectric resonant metaphotonics provides an alternative physical mechanism of light localization via low-order dipole and multipole Mie resonances that support highly efficient nanoresonators and all-dielectric metasurfaces~\cite{koshelev2020dielectric}. This so-called “Mie-tronics” approach~\cite{mie_won} successfully demonstrated the suppression of radiative losses of individual dielectric resonators while engineering unique optical modes with high-quality factors facilitating nonlinear effects at the nanoscale~\cite{Koshelev288}.\\
However, the refractive index $n$ attainable in the visible frequency range is relatively small (usually $n<5$), especially compared to the effective refractive index of metals or dielectrics at longer wavelengths. The currently available materials do not allow realizing a large manifold of concepts initially developed for plasmonics and microwave metamaterials, due to either large metallic losses or unavailability of materials with sufficiently large refractive index response at optical wavelengths. This issue presently confines conceptual demonstrations of many metaphotonics ideas at microwave frequencies~ \cite{shen2013planar,bonache2016controlling,yermakov2018experimental}.\\
Thus, a novel approach to break this limit is highly desirable for implementing many revolutionary ideas with a novel generation of low-loss optical materials and for applications of deep subwavelength optics. Here we introduce a novel platform to implement many of those ideas in lossless reflective materials that can help devise many applications. Our approach exploits an inverse design of deformed reflecting structures with user-defined nondispersive refractive indices and spatial form profiles. We discuss basic optical circuitry such as resonators with an ultra-high refractive index of $n=100$ and design a new class of hypersensitive sensors that showed a record value of sensitivity. We believe our approach will provide novel means of strong energy confinement, opening the door to integrated optoelectronic devices with simultaneous control of electrons and photons of large bandwidths in channels of comparable size.

\section*{Results}

\subsection*{Inverse-designed materials via geometrical deformations}

\begin{figure*}
  \centering
  \includegraphics[width=0.99\columnwidth]{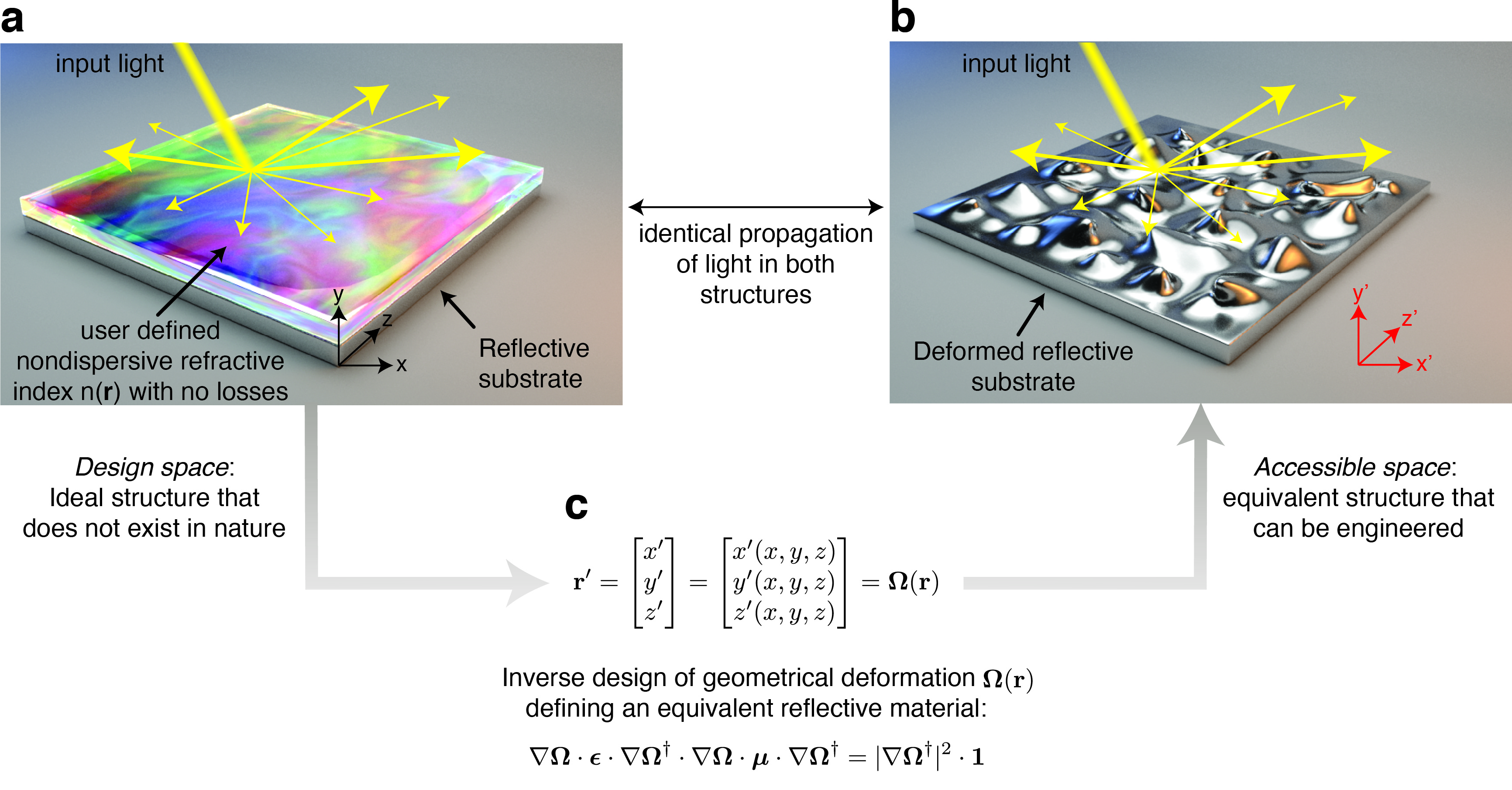}
  \caption{%
    \label{idea}
    \textbf{Designing materials via geometrical deformations: general principles.} \textbf{a}) Ideal structure for controlling light, characterized by a reflective substrate with a material possessing an arbitrary refractive index distribution on top. \textbf{b}) Equivalent deformed structure found by inverse design and composed of a reflective substrate in air that possesses the same electromagnetic behavior of (\textbf{a}). \textbf{c}) Mapping $\mathbf{r'}=\mathbf{\Omega}(\mathbf{r})$ function connecting both domains.
  }
\end{figure*}

The crucial challenge in controlling light localization at the nanoscale is the quantum mechanical structure of matter combined with reciprocity, representing the intuitive condition that a physical system cannot anticipate the future~\cite{doi:10.1002/9783527618156}. Quantum mechanics dictates that the susceptibility response $\chi(\omega)=n(\omega)^2-1$ of any accessible material is a frequency $\omega$ dependent function combining a series of Lorentz-type resonances~(see Chapter 3 of~\cite{10.5555/1817101}):
\begin{equation}
    \label{boy0}
    \chi(\omega)=\sum_m\left(\frac{a_m}{\omega_m-\omega}+\frac{a_m}{\omega_m^*+\omega}\right),
\end{equation}
with complex oscillator frequencies $\omega_m$ and amplitudes $a_m$. The causality principle manifests in Eq.~\eqref{boy0} as a rigid relationship, expressed by Kramers-Kronig (KK) relations~\cite{doi:10.1002/9783527618156}, between the real and imaginary part of $\chi(\omega)$. Due to the direct relationship between $\chi(\omega)$ and $n(\omega)$, KK relations state the impossibility to independently control the real part of $n(\omega)$, representing the effective refractive index of the material, from the material losses, defined by the imaginary part of the refractive index $n(\omega)$.\\
Metallic media possess atomic resonances $\omega_m$ at visible wavelengths~\cite{maier}, and acquire a strong light localization ability at visible frequencies. However, KK relations constraint high-localization frequencies to points of high absorption, blending the effects of energy localization and losses irreversibly. Dielectric materials, conversely, possess electronic resonances $\omega_m$ in the deep-blue or ultra-violet frequency range and exhibit transparency over large optical bandwidths. However, the accessible refractive index of these materials in the lossless window is inevitably modest, compromising the resulting localization power, especially if compared to plasmonic media.\\
Figure~\ref{idea} shows an alternative material platform that could overcome the issue mentioned above. The starting configuration is a universal basic structure composed by a reflective substrate lying in the $(x,z)$ plane, with a user-defined, semi-infinite material defined on top (Fig.~\ref{idea}a). The medium possesses a nondispersive and lossless anisotropic refractive index $\mathbf{n}(\mathbf{r})=\sqrt{\boldsymbol{\epsilon}(\mathbf{r})\cdot\boldsymbol{\mu}(\mathbf{r})}$, with $\boldsymbol{\epsilon}(\mathbf{r})$ and $\boldsymbol{\mu}(\mathbf{r})$ dielectric permittivity and magnetic permeability tensors, respectively, and $\mathbf{r}=(x,y,z)$ position coordinates.\\
While a material defined as such does not directly exist in nature, we here show that it is possible to engineer this structure by an inverse application of transformation optics~\cite{Pendry549,tra2,McCall_2018,So_2019}. Inverse design is a powerful method for creating a material with desired properties \cite{pilozzi2018machine,so2020deep} while transformation optics establishes a relationship between coordinate transformations and geometric materials equivalent to light propagation. Following this correspondence, we wish to implement the user-defined medium $\mathbf{n}(\mathbf{r})$ from a geometrical deformation of coordinates $\mathbf{r}'=\boldsymbol{\Omega}(\mathbf{r})$ with:
\begin{equation}
    \label{om0}
    \mathbf{r}'=\begin{bmatrix}
    x'\\y'\\z'\end{bmatrix}
    =\begin{bmatrix}
    x'(x,y,z)\\
    y'(x,y,z)\\
    z'(x,y,z)
    \end{bmatrix}=\boldsymbol{\Omega}(\mathbf{r}),
\end{equation}
with $\mathbf{r}'$ transformed coordinates. In the transformed space $\mathbf{r}'$, Maxwell equations remain identically the same with the introduction of a new material with dielectric permittivity $\boldsymbol{\epsilon}'(\mathbf{r}')=\frac{\nabla\boldsymbol\Omega\cdot\boldsymbol{\epsilon}\cdot
\nabla\boldsymbol\Omega^\dag}{|\nabla\boldsymbol\Omega^\dag|}$ and magnetic permeability $\boldsymbol{\mu}'(\mathbf{r}')=\frac{\nabla\boldsymbol\Omega\cdot\boldsymbol{\mu}\cdot\nabla\boldsymbol\Omega^\dag}{|\nabla\boldsymbol\Omega^\dag|}$ with $\nabla\boldsymbol \Omega=\frac{\partial\mathbf{r}'}{\partial\mathbf{r}}$ the Jacobian matrix~\cite{tra2}. 
Light cannot differentiate between the materials $\mathbf{n}(\mathbf{r})$ and  $\mathbf{n}'(\mathbf{r}')=\sqrt{\boldsymbol{\epsilon}'(\mathbf{r}')\cdot\boldsymbol{\mu}'(\mathbf{r}')}$, experiencing an identical evolution in the spaces $\mathbf r$ and $\mathbf r'$. 
For any response pair $n(\mathbf{r})$ and $n'(\mathbf r')$ created via $\boldsymbol \Omega(\mathbf{r})$, transformation optics guarantees the causality of the resulting material in both spaces $\mathbf r$ and $\mathbf r'$ (see Methods for a detailed demonstration).\\
The main idea is to inverse design the geometrical deformation \eqref{om0} so that the transformed refractive index is that of a vacuum, with $\boldsymbol{n}'(\mathbf{r}')=\mathbb{1}$: 
\begin{equation}
    \label{con0}
\nabla\boldsymbol\Omega\cdot\boldsymbol{\epsilon}\cdot\nabla\boldsymbol\Omega^\dag\cdot\nabla\boldsymbol\Omega\cdot\boldsymbol{\mu}\cdot\nabla\boldsymbol\Omega^\dag=|\nabla\boldsymbol\Omega^\dag|^2\cdot\mathbb{1}.
\end{equation}
When this special condition occurs, the transformed medium $\boldsymbol{n}'(\mathbf{r}')$ appears as a deformed reflector immersed in a vacuum, and the deformed surface defined from \eqref{om0} acquires the ability to emulate the material with the user-defined refractive index $\boldsymbol{n}$ (Fig. \ref{idea}b-c).\\ 
Equation \eqref{con0} comprises a system of nine nonlinear partial differential equations, to be solved for the unknown transformation $\boldsymbol{\Omega}(\mathbf{r})$ for a given desired material response in $\boldsymbol{\epsilon}(\mathbf{r})$ and $\boldsymbol{\mu}(\mathbf{r})$. As a boundary condition, we assume that for $\mathbf{r}\rightarrow\infty$ the transformation \eqref{om0} tends to the identity $\boldsymbol{\Omega}(\mathbf{r})=\mathbf{r}$, with $\nabla\boldsymbol\Omega=\mathbb{1}$. This condition ensures that Eq. \eqref{con0} represents physical materials localized in a finite area of space, while relaxing to a vacuum $\boldsymbol\epsilon(\mathbf{r})=\boldsymbol\mu(\mathbf{r})=\mathbf{n}(\mathbf{r})=\mathbb{1}$ in the far-field for $\mathbf{r}\rightarrow\infty$.\\ 
In this article, we begin to study the solution of Eqs.~\eqref{con0} for scalar dielectric or magnetic materials defined by a single quantity $\epsilon(\mathbf{r})\equiv\epsilon_{zz}(\mathbf{r})$ and $\mu(\mathbf{r})\equiv\mu_{zz}(\mathbf{r})$, assumed without loss of generality on the $z$ axis:
\begin{align}
    \label{emu2d0}
    &\boldsymbol{\epsilon}=\begin{bmatrix}
    \mathbb{1} &0\\
    0 &\epsilon(\mathbf{r})
    \end{bmatrix}, &\boldsymbol{\mu}=\begin{bmatrix}
    \mathbb{1} &0\\
    0 &\mu(\mathbf{r})
    \end{bmatrix}.
\end{align}
 The resulting refractive index $\mathbf{n}=\sqrt{\boldsymbol{\epsilon}\cdot\boldsymbol{\mu}}$ acquires a traditional expression depending on a single scalar quantity $n(\mathbf{r})=\sqrt{\epsilon(\mathbf{r})\mu(\mathbf{r})}$. Supplementary Note I shows that a necessary condition for the solution of Eq. \eqref{con0} is that the deformation \eqref{om0} is $z-$invariant with $\nabla=[\nabla_\perp,0]$ and $\nabla_\perp=[\frac{\partial}{\partial x},\frac{\partial}{\partial y}]$. In this condition, Eqs. \eqref{con0} reduce to:
\begin{equation}
    \label{2d0}
    \begin{bmatrix}
    \left(\nabla_\perp\Omega_\perp\cdot\nabla_\perp\Omega^\dag_\perp\right)^2 &0\\
    0 &n^2(\mathbf{r})
    \end{bmatrix}=|\nabla_\perp\Omega^\dag_\perp|^2\cdot\mathbb{1},
\end{equation}
with $\boldsymbol{\Omega}_\perp=[x'(x,y),y'(x,y)]$. Supplementary Note II shows that the solution to Eqs. \eqref{2d0} is an inverse conformal mapping~\cite{schinzinger1991conformal}, with $\boldsymbol{\Omega}=[\real(\Omega), \imag(\Omega)]$ defined from the real $\real$ and imaginary $\imag$ part of a single analytic function $\Omega(u)$ of complex coordinate $u=x+iy$ satisfying the scalar equation $n^2(\mathbf{r})=\left |\frac{d\Omega(u)}{du}\right |^2$. 
We solve this problem with a pseudospectral approach~\cite{fornberg_1996} based on suitably defined rational Chebyshev polynomials~\cite{doi:10.1002/nme.392} in the complex domain. We expand the unknown deformation $\Omega(u)$ as follows:
\begin{equation}
\label{cheb0}
    \Omega\left(\frac{u}{a}\right)=\frac{u}{a}+\beta_0+\sum_{m=1}^{\infty}\beta_m T_m\left(\frac{u-ia}{u+ia}\right),
\end{equation}
with $T_m(v)$ the Chebyshev polynomial of order $m$, $a$ an arbitrary spatial scaling constant, and $\beta_m$ unknown coefficients.
Chebyshev polynomials $T_m(v)$ are a complete basis that represents any complex function on the unit circle $-1\le |v|\le 1$~\cite{fornberg_1996}. In \eqref{cheb0}, we use a a Cayley transform $v=\frac{u-ia}{u+ia}$~\cite{kreyszig11} that maps the unit circle to the positive semi-infinite space $y\ge 0$ and $-\infty\le x\le \infty$, providing a rational series expansion that represents any analytic function $\Omega(u)$ in the complex domain. Additionally, we impose the condition $\beta_0=-\sum_{m=1}^{\infty}\beta_m$ to reduce Eq.~\eqref{cheb0} in the far-field to the identity transformation $\Omega(u/a)=u/a$, as the user can verify by direct substitution and as requested for the inverse solution of Eq. \eqref{con0}.\\
By expressing the derivative $\frac{dT_m(u)}{du}=m\cdot U_{m-1}(u)$ with Chebyshev polynomials of second kind $U_m$, we obtain the nonlinear equation for the inverse design of the transformation $\Omega(u)$: 
\begin{equation}
    \label{cheb1}
    n^2\left(\frac{\mathbf{r}}{a}\right)=\left| 1+ \frac{2 ia}{(u+ia)^2}\cdot \sum_{m=0}^{\infty}\beta_{m+1}(m+1)\cdot U_{m}\left(\frac{u-ia}{u+ia} \right) \right|^2.
\end{equation}
In the scalar limit of Eqs.~\eqref{emu2d0}, the inverse design problem is equivalent to finding the set of coefficients $\beta_m$ that satisfy Eqs.~\eqref{cheb1} for a user-defined refractive index $n(\mathbf{r})$ given at the input.
We solve Eq. \eqref{cheb1} by statistical learning via nonlinear regression~\cite{hastie_09_elements-of.statistical-learning}. Given a user-defined distribution of refractive index $n(\mathbf{r})$, we create a training dataset composed of a discrete number $n=0,1,...$ of refractive index values $n(\mathbf{r}_n)$ sampled on a Chebyshev grid $\mathbf{r}_n$~\cite{fornberg_1996}. Each dataset input-output couple $[n(\mathbf{r}_n),\mathbf{r}_n]$, when substituted in Eq.~\eqref{cheb1} originates a nonlinear equation for the regression coefficients $\beta_{m+1}$. These are calculated via nonlinear least-square, by using a trust-region convex minimization routine~\cite{ceres-solver}. After solving for the coefficients $\beta_{m+1}$, Eq. \eqref{cheb0} predicts the deformation of coordinates $\Omega(u)$ for any point $\mathbf{r}'$ of the space and completes the solution of the problem.\\
In the family of materials designed via \eqref{cheb1}, the refractive index $n(\mathbf{r})$ is solely defined from the spatial curvature of $\Omega(u)$ arising from $\left |\frac{d\Omega(u)}{du}\right |^2$. As such, $n(\mathbf{r})$ does not show any theoretical limit between zero and infinity, with accessible values limited to implementing the required curvature arising from the solution of \eqref{cheb1}.

\subsection*{Nanoresonators with ultra-high refractive index}
\begin{figure*}[htbp]
  \centering
  \includegraphics[width=1.0\columnwidth]{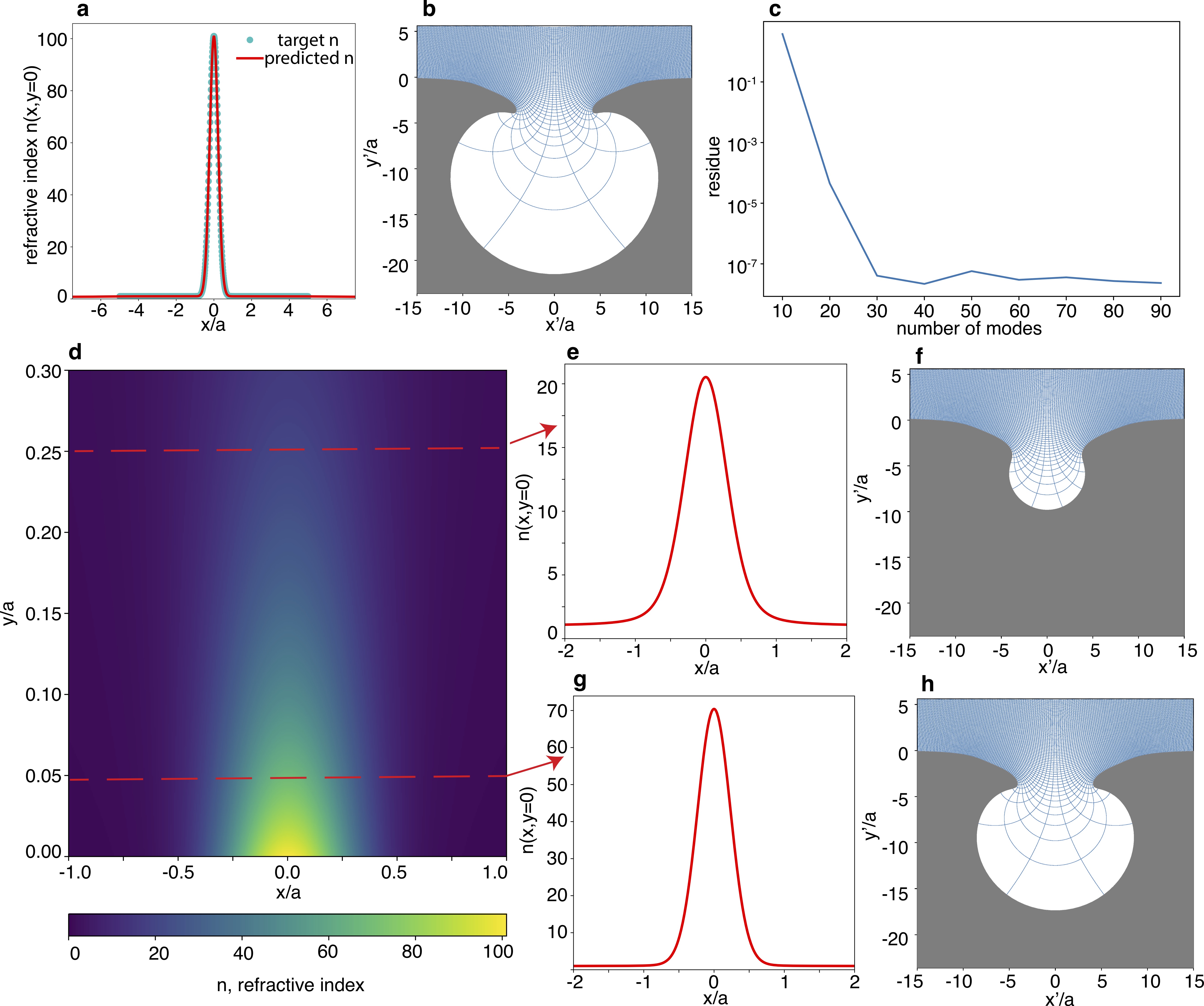}
  \caption{%
    \label{pcolor_panel}
    \textbf{Ultra-high refractive index nanoresonators.} \textbf{a}) Target Refractive index profile (circle markers) and inverse design refractive index (solid red line). \textbf{b}) Corresponding deformation in the accesible space. \textbf{c}) Convergence of nonlinear regression for a varying number of modes $M$. \textbf{d}) Spatial distribution of inverse designed refractive index in the design space. \textbf{e}-\textbf{g}) Refractive index distribution with amplitudes $n=20$ and $n=70$, respectively, and \textbf{f}-\textbf{h} corresponding deformations.
  }
\end{figure*}

At visible frequencies, high refractive index available materials with $n\le 5$ are typically III-IV semiconductors including Ge, Si, GaAs, GaSb. In these media, it is traditionally challenging to exploit their highest values of the refractive index because of high losses~\cite{kasap2017springer,baranov2017all}, and applications typically rely on $3\le n\le 4$~\cite{miroshnichenko2015nonradiating,Koshelev288,Liu2016AOM,Schmid_2017,YANG20171,PhysRevLett.119.243901,fran, melik2021fano,yurirev,zhizhchenko2019single}. To the best of the authors' knowledge, the reported highest refractive index in the visible is $n=5.1$ for a block copolymer self-assembly metasurface~\cite{kim2016highly}.\\
We here illustrate how it is possible via inverse design to engineer materials with ultra-high refractive index $(n\approx 100)$ and user-defined profile. We set a Gaussian distribution of refractive index $n(\mathbf{r})$ with amplitude $n(\mathbf{r}=0)=100$ and planar width $x/a=0.1$. Figure~\ref{pcolor_panel}a shows the corresponding training dataset (gray circle markers). In the solution of \eqref{cheb0}, we evaluate the minimum number $M$ of coefficients $\beta_{m+1}$ required to inverse solve the problem by increasing $M$ and computing at each stage the prediction mean square error $MSE=||n(\mathbf{r})-n_{pred}(\mathbf{r})||$, with $n_{pred}(\mathbf{r})$ the refractive index predicted from the transformation $\Omega(u)$. Few modes with $M=30$ yield a solution with accuracy $10^{-7}$ (Fig.~\ref{pcolor_panel}b ), and prediction values along the axis $x$ matching the required Gaussian profile (Fig.~\ref{pcolor_panel}a solid red line). Figure~\ref{pcolor_panel}c illustrates the refractive index distribution in the design space $\mathbf{r}$. The index relaxes exponentially in the $y-$axis, providing a diffusive-like nanoresonator profile with an ultra-high refractive index region defined in a narrow length along $y$.\\
Figure~\ref{pcolor_panel}d (solid gray area) shows the reflective deformation in the accessible space $\mathbf{r}'$ that emulates the material in Fig.~\ref{pcolor_panel}c. The deformation corresponds to the mapping $\mathbf{r'}=\boldsymbol\Omega(\mathbf{r})$ for $y=0$ and $-\infty<x<\infty$. The solid blue lines in Fig.~\ref{pcolor_panel}d report the coordinate deformation in the remaining area of space for $y\neq 0$.
A single coordinate transformation $\Omega(u)$ defines an entire family of materials with different distributions of refractive indices. If we consider a coordinate line at varying $x$ and constant $y_0$ in Fig.~\ref{pcolor_panel}c, and set $y_0$ as a new origin of coordinate along $y$ via $y\rightarrow y-y_0$, this deformation generates a new reflective material in the accessible space $\mathbf{r}$ with the index distribution defined by $n(\mathbf{r})$ evaluated at the new origin $y=0$ and varying $x$. The corresponding refractive index maintains the same spatial profile along $x$, but acquires a different amplitude due to the relaxation of the index values towards infinity (Fig.~\ref{pcolor_panel}c). Figures~\ref{pcolor_panel}e-h illustrate this possibility, showing the engineering of different reflective deformations yielding Gaussian index profiles with maxima of $n=20$ and $n=70$.\\
Supplementary Fig. 1 illustrate examples of ultra-high index resonators realized with different form factors. Panels a-c show the case of a flat-top resonator with a super-Gaussian spatial index profile. Supplementary Fig. 1d-f shows the example of two adjacent resonators with different refractive indices, demonstrating the flexibility of this technique in designing complex index modulations that are not possible to observe at visible frequencies with conventional materials.

\subsection*{Resonator modes}
\begin{figure*}[htbp]
  \centering
  \includegraphics[width=1.0\columnwidth]{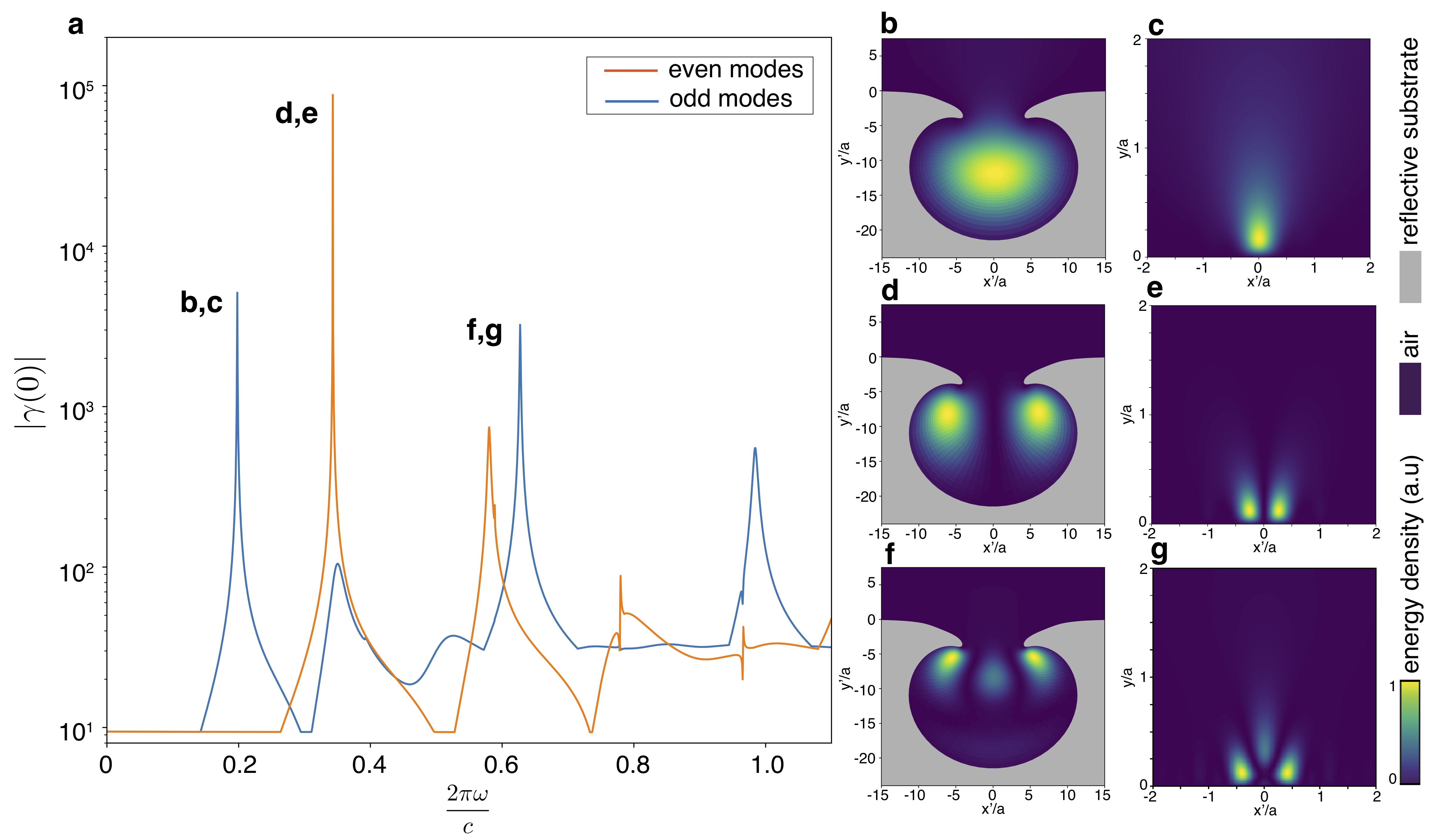}
  \caption{%
    \label{panel3.5}
    \textbf{BIC states of inverse designed ultra-high index resonators.}
    \textbf{a}) Eigenvalue amplitude $|\gamma(0)|$ vs. frequency $k$. \textbf{b})-\textbf{g}) Electromagnetic energy spatial distribution for singular states in the accessible $\mathbf{r'}$ and design $\mathbf{r}$ spaces, respectively.
      }
\end{figure*}

Resonators designed by complex geometrical deformations support singular energy localization states appearing as Bound States in the Continuum (BIC)~\cite{BogdanovFratalocchiKivshar}. We studied singular resonances from the user-defined space $\mathbf{r}$, by decomposing the electric field $\mathbf E=\mathbf E_t+\mathbf E_y$ into a transverse $\mathbf E_t$ component, lying on the $(x,z)$ plane, and a normal contribution $\mathbf E_y$ along $y$. We then define the following eigenvalue problem for the transverse field:
\begin{equation}
    \label{eigv0}
    \frac{\partial }{\partial y}\mathbf{E}_t(\mathbf r)=\gamma(y)\cdot\mathbf{E}_t(\mathbf r),
\end{equation}
defining the spectral decomposition of the operator $\frac{\partial}{\partial y}$ along the normal $y$ axis. Of particular interest are trapped modes arising from the singularities in the eigenvalue amplitude $|\gamma(0)|\rightarrow\infty$ arising at $y=0$. For the corresponding eigenvector $\mathbf{E}_t(\mathbf r)$ to be bounded and physical, the contribution $\frac{\partial }{\partial y}\mathbf E_t$ is required to stay finite. This condition implies that $\mathbf{E}_t\rightarrow 0$ at $y=0$, satisfying the boundary condition $\mathbf E_t=0$ imposed by a perfect reflector over the entire $(x,z)$ plane for transverse-electric (TE) polarized modes. Each singularity of Eq.~\eqref{eigv0} therefore defines a TE-polarized, nonradiating state that exists without any impinging source, by autonomously satisfying the boundary conditions on the reflector plane at $y=0$. The corresponding eigenvector $\mathbf E_t(\mathbf r)$ furnishes the spatial profile of the resulting TE-polarized BIC.
As the singularity originating the BIC arises at $y=0$, the electromagnetic field composing the BIC mode typically localizes in the proximity of the reflective plane at $y=0$. In the accessible space $\mathbf{r}'$, such energy localizations appear in the vacuum-gaps areas created by geometric mapping $\boldsymbol\Omega(\mathbf{r})$ (Fig.~\ref{idea}b). The case for TM polarization follows the same analysis, with the only difference of decomposing the magnetic field $\mathbf H=\mathbf H_t+\mathbf H_y$ into transverse $\mathbf H_t$ and normal $\mathbf H_y$ contributions, and formulating Eq.~\eqref{eigv0} for $\mathbf H_t$.
The boundary condition of TM-polarized localizations, $\mathbf H_t=0$ at $y=0$, implies that the energy maxima of these modes localize on the reflector surface. TE localizations, conversely, possess an electric field that completely localizes in the vacuum. As the latter configuration is more advantageous for the applications discussed in this work, we focus on TE modes while deferring the analysis of TM-polarized BIC to future work.\\ 
We illustrate these results by solving the eigenvalue Eq.~\eqref{eigv0} with an approach that generalizes the work of~\cite{PAGNEUX20101834}. The idea is to project the eigenvalue equation over a basis of scattering modes computed at $\mathbf{r}\rightarrow\infty$, and resolve the resulting discrete equations by a spectral method. Supplementary section III discusses implementation details of this approach.
Figure \ref{panel3.5} illustrates calculation examples of BIC supported by the ultra-high refractive index nanoresonator designed in Fig.~\ref{pcolor_panel}a. Figure \ref{panel3.5}a plots the corresponding eigenvalue amplitude $|\gamma (0)|$ at $y=0$ for a varying frequency measured in wavevector units $k=2\pi\omega/c$. As the resonator structure is symmetric in the $(x,z)$ plane, we perform computations for even (solid blue lines) and odd (solid orange lines) BIC modes separately. The eigenvalue $\gamma$ shows the formation of clear singular states with amplitudes reaching up to $10^5$ at $k=0.3429$. Each of these singularities defines a BIC state with a different energy profile. Figure \ref{panel3.5}b reports the spatial energy distribution in the accessible space, while panel c shows the corresponding profile in the design space. In the design space, odd and even TE polarized BIC modes confine electromagnetic energy inside the ultra-high refractive region in proximity to the reflective surface. In the accessible space, these modes manifest as narrow light localization in the vacuum gaps defined by the complex reflector spatial profile.\\
An advantage of this form of energy localization is the possibility to access the entire spatial distribution of high refractive index, as observed from the mapping between the design and accessible spaces in Fig.~\ref{panel3.5}b-c. This property results from the user-defined refractive index region created by the surface profile of the deformed reflector, which provides trapping areas for singular states with a high degree of energy localization. In the next Section, we leverage this feature to design a new family of compact hypersensitive devices.     

\subsection*{Sensing application}
\begin{figure*}[htbp]
  \centering
  \includegraphics[width=1.0\columnwidth]{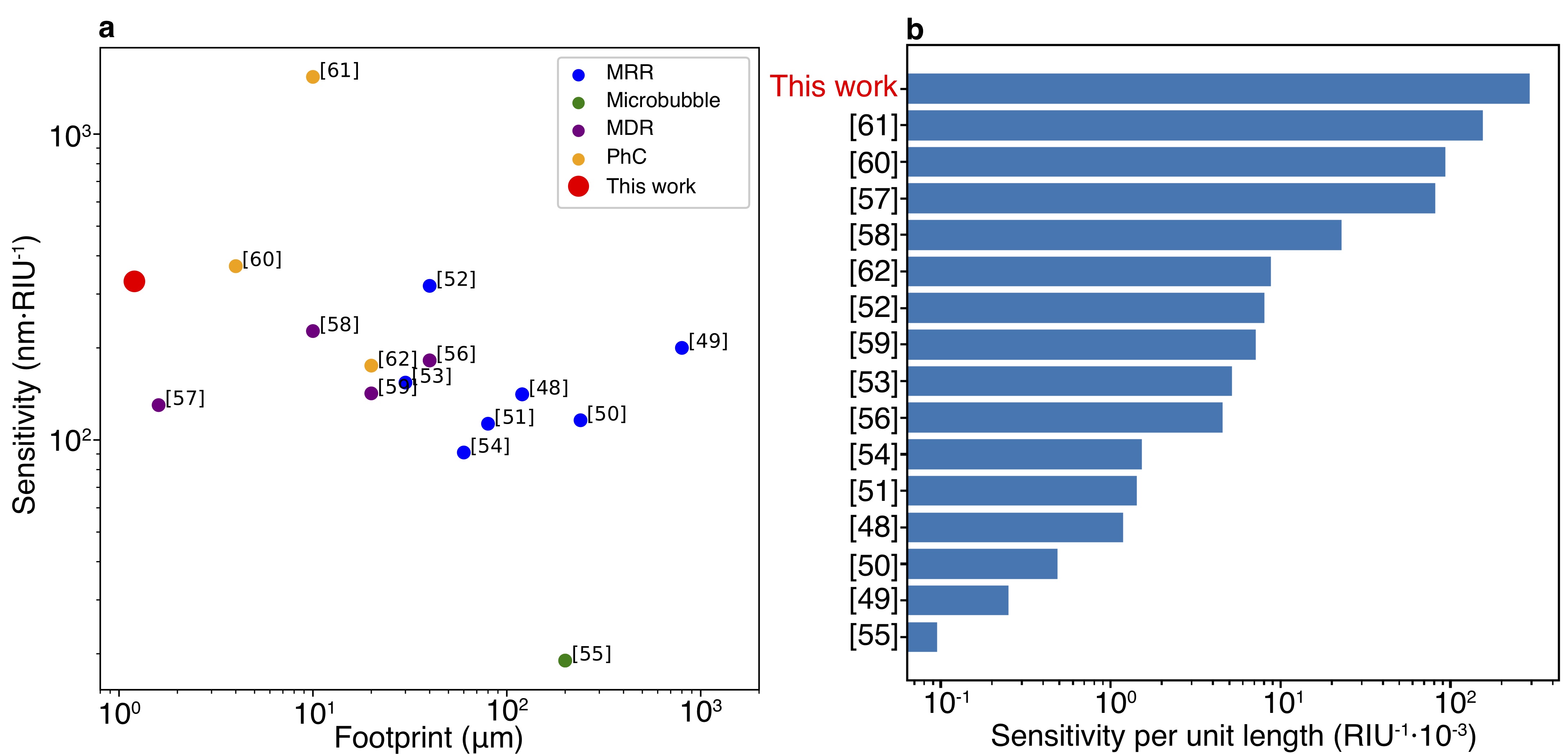}
  \caption{%
    \label{sensor_fom}
    \textbf{State-of-the-art in RI optical sensors.} \textbf{a}) Sensitivity vs. sensor footprint. \textbf{b})  Single figure of merit displaying the sensitivity per sensor footprint.
  }
\end{figure*}

\begin{figure*}[htbp]
  \centering
  \includegraphics[width=1.0\columnwidth]{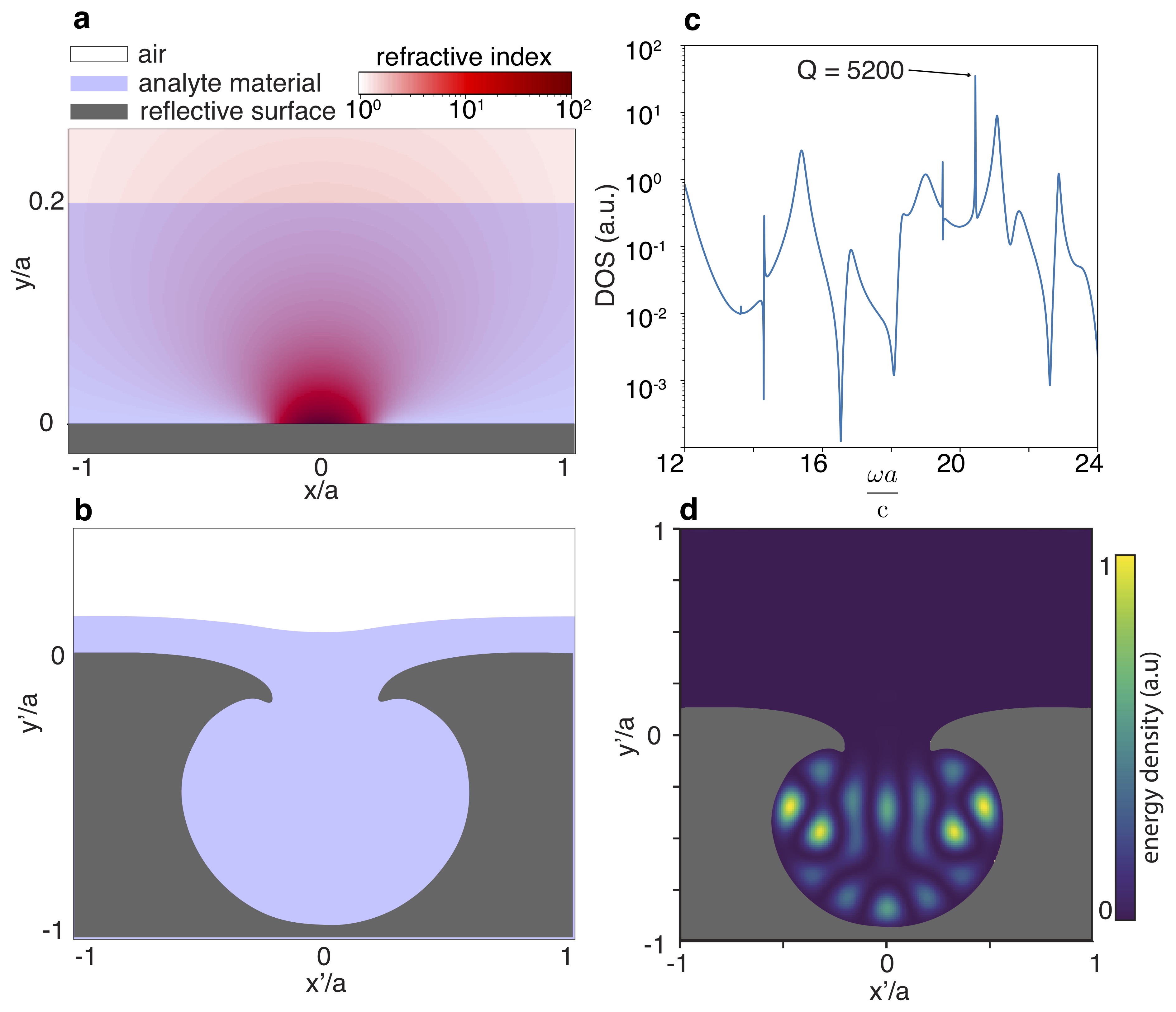}
  \caption{%
    \label{ldos}
    \textbf{RI sensor based on inversely designed ultra-high index material}
    \textbf{a}) Sensor idea in the design space comprising a single resonator with ultra-high refractive index of $n=100$ and an analyte placed on top. \textbf{b}) Corresponding configuration in the accessible space. \textbf{c}) Density of states (DOS) of the resonator (solid blue line) with BIC states possessing $Q=5200$). \textbf{d}) Electromagnetic energy spatial distribution of the BIC state in the accessible space.
  }
\end{figure*}

\begin{figure*}[htbp]
  \centering
  \includegraphics[width=1.0\columnwidth]{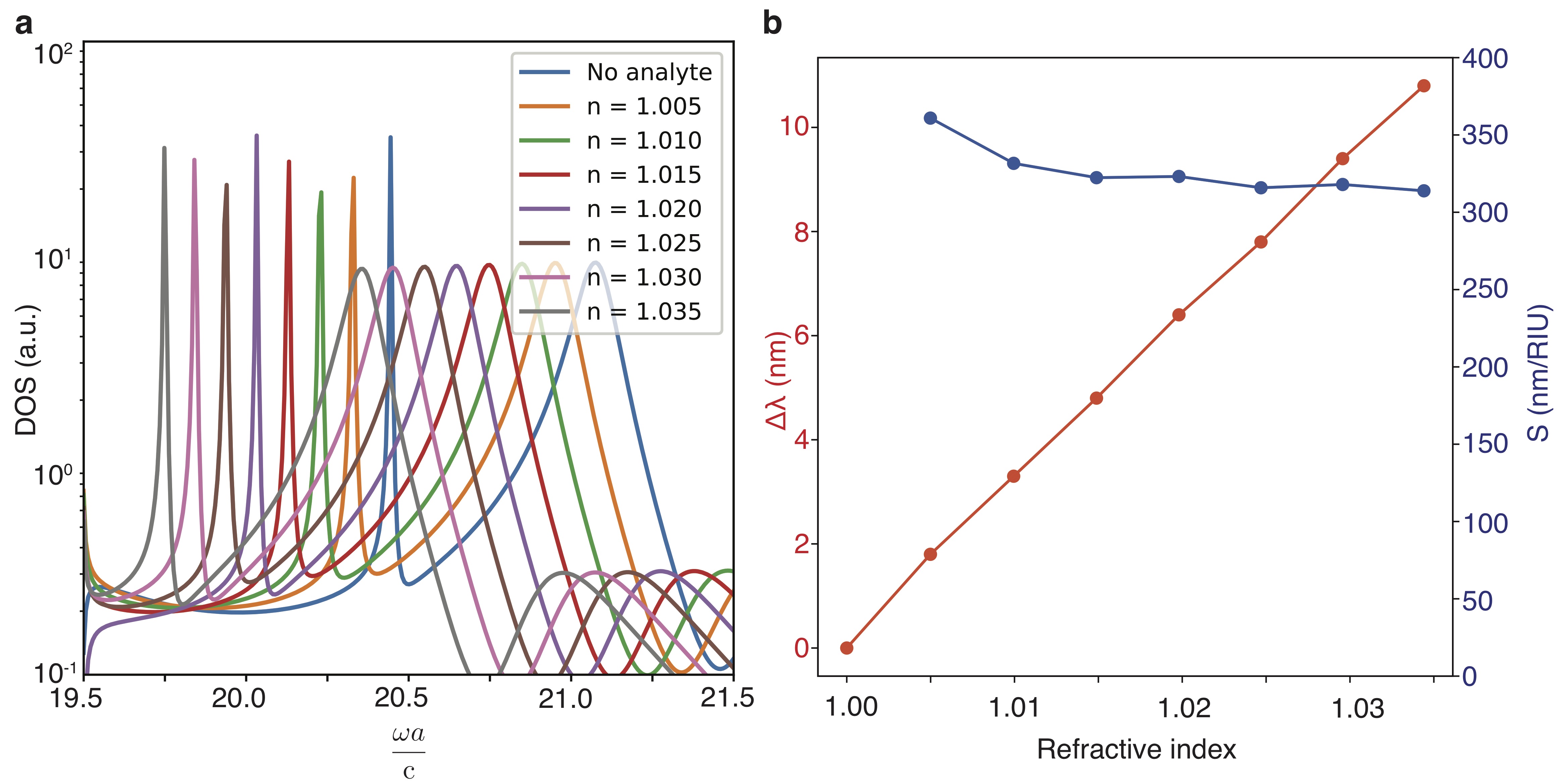}
  \caption{%
    \label{sensor_res}
    \textbf{Performance of the RI sensing device.} \textbf{a}) Spectral shift of the BIC resonant wavelength for a varying refractive index $n$ of the analyte. \textbf{b}) Wavelength shift $\Delta \lambda$ and differential sensitivity $S$ vs. refractive index change of the analyte
  }
\end{figure*}

Figure~\ref{sensor_fom}a summarizes the state-of-the-art in refractive index (RI) sensing in sensitivity, measured in nm/RIU, vs. sensor footprint, reported in units of microns \cite{yalcin2006optical,kim2008integrated,girault2015integrated,schmidt2014improving,liu2016simultaneous,wei2016direct,goykhman2010ultrathin,tang2016packaged,hu2008planar,wang2013suspended,kim2019chip,grist2013silicon,hamed2013photonic,di2009chemical,garcia2010label}. Figure~\ref{sensor_fom}b order these results in terms of a single FOM measuring the differential sensitivity per sensor pathlength, obtained as the ratio between the differential sensitivity and the sensor footprint. Sensors created with single resonator structures have typically lower performances than arrays of periodic structures such as, e.g., photonic crystals (PhC), which can reach sensitivities up to 1500 nm/RIU. Addressing sensor miniaturization without compromising performances is challenging: single resonator structures cannot benefit from nearest-neighbor constructive interactions and leverage only the quality-$Q$ factor of the resonant modes supported by the refractive index of the resonator. While in microdisk resonators with footprints of few microns, the $Q$ factor can reach $Q=12000$ \cite{wang2013suspended}, this value decreases to $Q=680$ in the presence of analytes, compromising the overall performances. A detrimental issue is also the spatial distribution of resonator modes, which usually localize their energy in high refractive index regions. These areas are the portion of resonator space inaccessible from the environment and that analytes cannot probe.
As a result of these shortcomings, sensitivities up to $300$~nm/RIU requires single resonators with large footprints of tens of microns.\\
The inverse design platform developed in this work can help to address these issues by designing compact resonators with ultra-high refractive index profiles and exploiting the high-quality BIC resonances supported by these structures. Figure~\ref{ldos}a-b illustrates the main design idea. We consider a single resonator in the design space possessing an ultra-high refractive index of $n=100$ and Gaussian form factor, as shown in Fig.~\ref{pcolor_panel}c, and place an analyte on top (Fig.~\ref{ldos}a blue area). In this configuration, the analyte permeates the entire refractive index region, including the areas of $n=100$. In the accessible space (Fig.~\ref{ldos}b), the structure appears as a deformed reflector with a conformal layer of material that fills up the cap space created by the modulated surface.\\
Figure~\ref{ldos}c presents the density of states (DOS) of the resonator. We perform this calculation with Finite-Difference Time-Domain (FDTD) simulations. We extract quality $Q$ factors via a robust Prony method obtained by augmenting the original Prony scheme with statistical regression (see details in Methods). The structure supports diverse BIC states with quality factors up to $Q=5200$, which we choose for our analysis. By setting the unit length $a=1~\mu$m, the BIC with the highest quality factor occurs at the resonant wavelength is $\lambda_{BIC}=307~$nm, in the ultra-violet (UV) region. Figure~\ref{ldos}d illustrates the spatial energy distribution of the BIC in the accessible space. The BIC provides robust electromagnetic energy localization inside the deformed reflector, showing a complex distribution of multiple energy hot spots. In this sensor configuration, the analyte fills up the entire gap area (Fig.~\ref{ldos}b) and experiences a strong light-matter interaction by probing the high energy spots supported by the BIC state.\\
Figure~\ref{sensor_res} analyzes this mechanism in detail by computing the differential sensitivity from the wavelength shift of the BIC resonance. We computed these results from FDTD simulation by changing the refractive index of the analyte with steps $\Delta n=5\cdot 10^{-3}$~RIU. Figure~\ref{sensor_res}a reports the variation of the density of states for analytes with refractive indices between $n=1.005$ and $n=1.035$. While we do not observe any appreciable variation of the $Q-$factor, the system shows a perfect linear shift of the BIC resonant frequency (Fig.~\ref{sensor_res}b solid red line). The resulting differential sensitivity stays between $S=350$~nm/RIU and $S=310$~nm/RIU, with an average value of $S=330$~nm/RIU  (Fig.~\ref{sensor_res}b solid blue line). This design reports a record value of sensitivity per unit pathlength (Fig.~\ref{sensor_fom}b red circle), providing a single resonant structure with similar performances of periodic assembly of resonators with footprints of one order magnitude larger (Fig.~\ref{sensor_fom}a).

\section*{Conclusion}
We have presented an inverse-design framework for implementing structures with user-defined nondispersive refractive indices and spatial form profiles in suitably deformed reflectors. We have discussed the basic optical circuitry, such as resonators with an ultra-high refractive index of $n=100$ in diverse user-defined shapes. Leveraging the high refractive index response of these systems, we have designed a new class of hypersensitive sensors that demonstrate record values of differential sensitivity per unit pathlength $S_{pl}=290 $ nm$\cdot\mu$m$^{-1}\cdot\mathrm{RIU}^{-1}$.\\
Notably, such inverse metaphotonic reflectors can overcome two challenging problems. They allow (i) to design structures with similar trapping performances of plasmonic structures but with no losses and (ii) to implement dielectric materials with user-defined refractive indices that go beyond the values available in physical materials. While these inverse-designed reflectors do not eliminate all issues, they replace physical limitations arising from the points (i) and (ii) with a more straightforward problem of engineering a surface with a user-implemented profile. Some existing nanofabrication techniques are already relevant to address these issues. While an exhaustive analysis is beyond the scope of this conceptual work, here we can mention two-photon lithography, which can already reach resolutions in below the micron range \cite{geng2019ultrafast} and is successively covered by a reflective layer, or powder bed fusion, which is essentially a version of three-dimensional printing for metallic structures~\cite{elder2020nanomaterial}.

\section*{Acknowledgements}
The authors acknowledge the use of resources of the Supercomputing Laboratory at KAUST. Y.K. acknowledges a support from the Australian Research Council (grant DP210101292), Russian Science Foundation (grant 21-72-30018), and the US Army International Office (grant FA520921P0034).

\section*{Methods}
{\bf Causality}. We here prove that a material arising from the solution of Eq. \eqref{con0} is physical and obeys the causality principle. Causality in optics is observed via Kramers-Kronig relationships~\cite{doi:10.1002/9783527618156}, which relate the real and imaginary part of the dielectric susceptibility $\boldsymbol{\epsilon}(\mathbf{r},\omega)=\boldsymbol{\epsilon}_r(\mathbf{r},\omega)+i\boldsymbol{\epsilon}_i(\mathbf{r},\omega)$ and magnetic permeability $\boldsymbol{\mu}(\mathbf{r},\omega)=\boldsymbol{\mu}_r(\mathbf{r},\omega)+i\boldsymbol{\mu}_i(\mathbf{r},\omega)$ expressed in the frequency domain~$\omega$:
\begin{align}
    \label{kk0}
    \boldsymbol{\sigma}(\mathbf{r},\omega)=\frac{1}{i\pi}\mathrm{PV}\int_{\infty}^{\infty}\frac{\boldsymbol{\sigma}(\mathbf{r},\omega)}{\omega'-\omega}d\omega',
\end{align}
with $\boldsymbol{\sigma}(\mathbf{r},\omega)=\boldsymbol{\epsilon}(\mathbf{r},\omega)$ or $\boldsymbol{\sigma}(\mathbf{r},\omega)=\boldsymbol{\mu}(\mathbf{r},\omega)$, and $\mathrm{PV}$ indicating the principal value. 
Multiplying Eq. \eqref{kk0} by $\frac{\nabla\Omega^\dag}{\left|\nabla\Omega^\dag\right|^{\frac{1}{2}}}$ from the right and by $\frac{\nabla\Omega}{\left|\nabla\Omega\right|^{\frac{1}{2}}}$ from the left, it becomes:
\begin{align}
    \label{kk1}
    &\boldsymbol{\sigma}'(\mathbf{r}',\omega)=\frac{1}{i\pi}\mathrm{PV}\int_{\infty}^{\infty}\frac{\nabla\Omega(\mathbf{r})\cdot\boldsymbol{\sigma}(\mathbf{r},\omega)\cdot\nabla\Omega^\dag(\mathbf{r})}{|\nabla\Omega^\dag(\mathbf{r})|\cdot\left(\omega'-\omega\right)}d\omega',
\end{align}
which represents Kramers-Kronig relationships for the optical transformed material properties $\boldsymbol{\sigma}'(\mathbf{r}',\omega)=\boldsymbol{\epsilon}'(\mathbf{r}',\omega),\boldsymbol{\mu}(\mathbf{r}',\omega)$ in the transformed physical coordinates $\mathbf{r}'=\boldsymbol{\Omega}(\mathbf{r})$.
As discussed in the previous section, the optical permittivity $\boldsymbol{\epsilon}=\boldsymbol{\epsilon}(\mathbf{r})$ and permeability $\boldsymbol{\mu}=\boldsymbol{\mu}(\mathbf{r})$ emulated via Eq. \eqref{con0} are nondispersive and depend only on the spatial coordinates $\mathbf{r}$. Under this condition, the causality condition Eq. \eqref{kk1}:
\begin{align}
    \label{kk2}
    &\boldsymbol{\sigma}'(\mathbf{r}')=\frac{\nabla\Omega(\mathbf{r})\cdot\boldsymbol{\sigma}(\mathbf{r})\cdot\nabla\Omega^\dag(\mathbf{r})}{|\nabla\Omega^\dag(\mathbf{r})|}\cdot\mathrm{PV}\int_{\infty}^{\infty}\frac{d\omega'}{i\pi(\omega'-\omega)}=\frac{\nabla\Omega(\mathbf{r})\cdot\boldsymbol{\sigma}(\mathbf{r})\cdot\nabla\Omega^\dag(\mathbf{r})}{|\nabla\Omega^\dag(\mathbf{r})|},
\end{align}
reduces to transformation optics relationships of the  material response, demonstrating that a material designed with this approach is perfectly causal. This result relies on the geometrical nature of transformation $\boldsymbol{\Omega}(\mathbf{r})$ used to create the material in \eqref{con0}, which does not depend on the frequency $\omega$. In this case, $\mathrm{PV}\int_{\infty}^{\infty}\frac{1}{i\pi(\omega'-\omega)}d\omega'=1$ and the real and imaginary part of Eq. \eqref{kk2} are decoupled, allowing to engineer a material that is fully causal.

{\bf BIC Q-factor computations}. We perform numerical simulations with our parallel NANOCPP FDTD solver. In the FDTD analysis, we studied the response of the structure on broadband TE-polarized light from the TFSF source by probing the electric field inside the nanoscale resonator and then by constructing the density of states (DOS). To resolve the $Q$-factor of high-Q modes within the DOS, we analyze the mode's energy decay by augmenting the Prony algorithm with a  cross-validation model, as explained in the following. The Prony algorithm decomposes the signal into a sum of complex exponential:
\begin{equation}
    \label{mode_decom}
\hat{f}(t) = \sum_{i=1}^{M} A_i e^{-\sigma_i t} \cos(\omega_i t + \phi_i)
\end{equation}
with amplitudes $A_i$, decaying constants $\sigma_i$, frequencies $\omega_i$ and phases $\phi_i$. To compute these terms, we first divide the FDTD computed over time into two distinct sets: a training set from $t=t_0$ to $t=t_1$, and a testing set from $t=t_1$ to the simulation time end $t=t_2$. We chose the intervals $t_1-t_0$ and $t_2-t_1$ to be approximately equal.\\
For a given input number of modes and sampling rate, we first solve linear prediction model for the training set with least-squares algorithm, find roots of a characteristic polynomial of the model, and then solve the original set of linear equations to obtain amplitudes and phases of an individual mode. Next, we apply this solution to the testing test and calculate the mean squared error between the FDTD data and the Prony model prediction. We then construct a surface of MSE values for different values of number of modes $M$ and sampling rates $T_s$, and select the  values of $M$ and $T_s$ that yield the best prediction. For each mode, we then compute the Q-factors as:
\begin{equation}
    \label{Q-fac}
Q=\frac{\omega_i}{2\sigma_i}
\end{equation}

\newpage

\clearpage

\clearpage

\end{document}